\documentclass{article}
\usepackage{spconf,amsmath,graphicx}
\usepackage{cite}
\usepackage{hyperref}
\hypersetup{
    colorlinks=true,
    linkcolor=blue,
    filecolor=magenta,      
    urlcolor=cyan,
    pdftitle={Overleaf Example},
    pdfpagemode=FullScreen,
    }
\usepackage{pdfpages}
\usepackage{booktabs}
\usepackage{multirow}
\usepackage{float} 
\usepackage{amsfonts,amssymb}
\usepackage[caption=false]{subfig}
\usepackage{color}

\title{Hierarchical Metadata Information Constrained Self-Supervised Learning for Anomalous Sound Detection Under Domain Shift}

\name{Haiyan Lan$^{1}$, Qiaoxi Zhu$^2$, Jian Guan$^{1*}$\thanks{*Corresponding Author}, Yuming Wei$^1$, Wenwu Wang$^3$
\thanks{This work was partly supported by the Natural Science Foundation of Heilongjiang Province under Grant No. LH2022F010 and No. YQ2020F010, and a Newton Institutional Links Award from the British Council with Grant No. 623805725.
}}
\address{
  $^1$Group of Intelligent Signal Processing, College of Computer Science and Technology,\\  Harbin Engineering University, China\\
  $^2$Centre for Audio, Acoustics and Vibration, University of Technology Sydney, Australia\\
  $^3$Centre for Vision Speech and Signal Processing, University of Surrey, UK}
\begin{document}
%
\maketitle
\begin{abstract}

Self-supervised learning methods have achieved promising performance for anomalous sound detection (ASD) under domain shift by incorporating the metadata of domain shift types and machine sound attributes in feature learning. However, the relation between domain shifts and machine sound attributes has yet to be fully utilised despite their potential benefits for better characterising domain shifts. This paper presents a hierarchical metadata information constrained self-supervised ASD method, where the hierarchical relation between domain shift types (section IDs) and attributes is constructed and used as constraints to improve feature representation. In addition, we propose an attribute-group-centre-based method for calculating the anomaly score under the domain shift condition. Experiments show improved audio feature learning over the state-of-the-art methods in DCASE 2022 challenge Task 2. 

\end{abstract}
\section{Introduction}
\label{sec:intro}
Anomalous sound detection (ASD) is a task for automatically identifying the working condition of a machine as normal or abnormal based on the sound emitted from the machine. Due to the difficulty in collecting rare and diverse anomalous sounds, it is a challenging unsupervised learning task with only normal sounds available for model training \cite{DCASE2020}. Unsupervised ASD methods based on autoencoder (AE) \cite{IDNN, GMAE, IDCAE} or self-supervised classification models \cite{Selfsupervised, cnnfeature} with metadata (e.g. machine IDs) incorporated achieved state-of-the-art performance on the Detection and Classification of Acoustic Scenes and Events (DCASE) challenge 2020 Task 2 \cite{DCASE2020}. However, ASD often has limited performance in practice due to the domain shift problem \cite{baseline22}. That is, acoustic characteristics differ between the source domain (in training) and the target domain (in detection), with the change of attributes \cite{baseline22} (e.g., machine operating conditions or types of noise). Due to this problem, the anomalies in the target domain can be misidentified with the model trained using sounds from the source domain.

\begin{figure*}[htbp]
    \centering
        \subfloat[Training procedure of the proposed model]{
             \label{fig:model}
             \includegraphics[width=1.2\columnwidth]{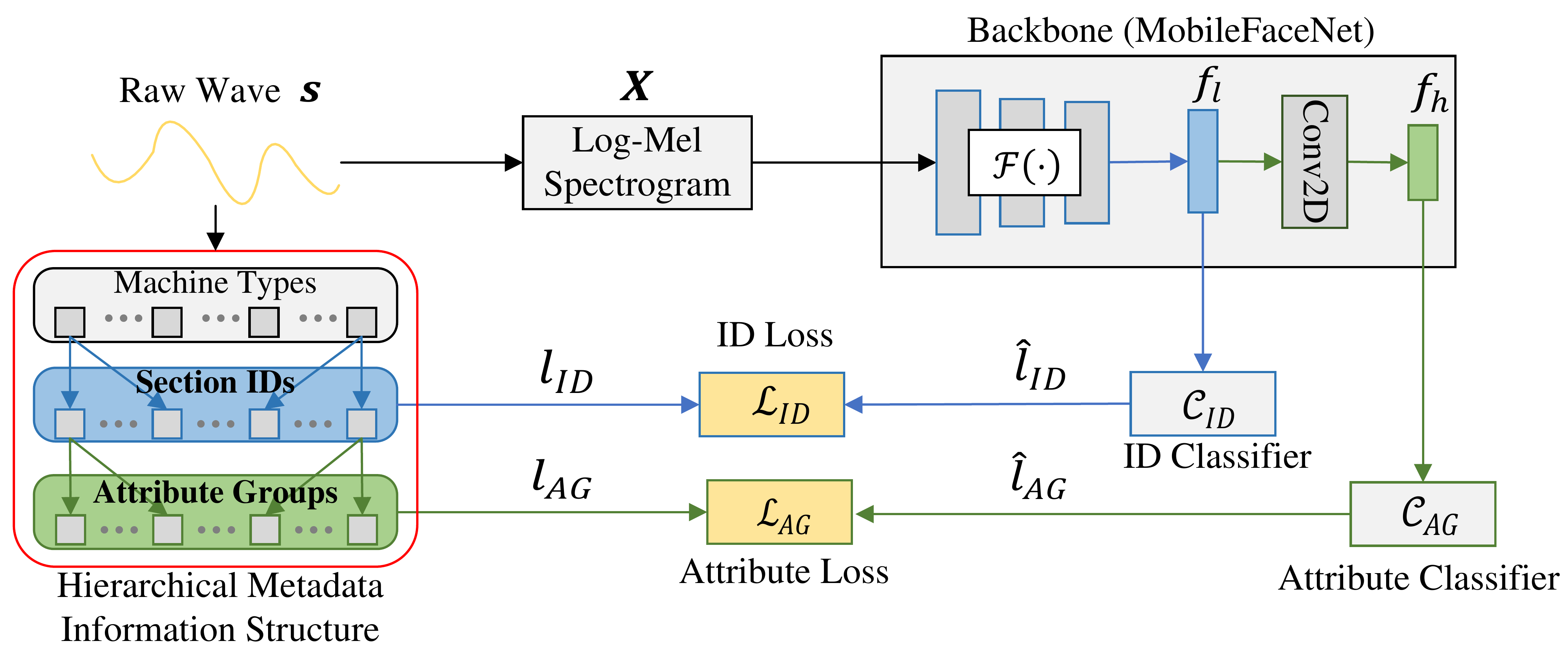}
             }
        \subfloat[Hierarchical metadata information structure]{
             \label{fig:metadata}
             \includegraphics[width=0.64\columnwidth]{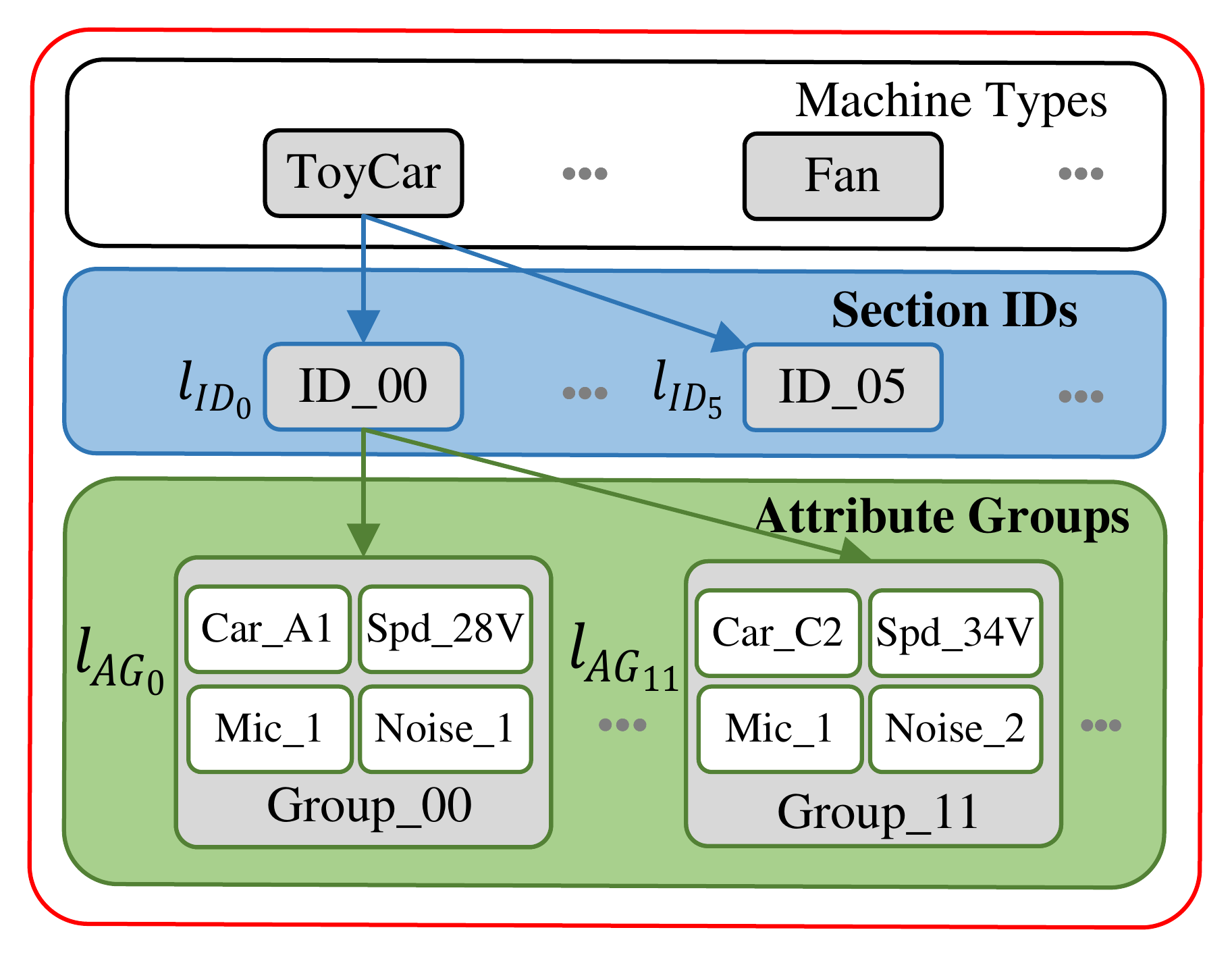}
             }
        \label{fig:framework}
        \caption{Framework of the proposed HMIC method, where the hierarchical relation of the metadata information is exploited by the introduced hierarchical metadata information structure, and a backbone network (i.e., MobileFaceNet \cite{MFN}) is used for the extraction of low-level feature $\textbf{\textit{f}}_{l}$ and high-level feature $\textbf{\textit{f}}_{h}$, which are constrained by section ID label $l_{ID}$, and attribute group label $l_{AG}$, respectively. Here, $\mathcal{F}(\cdot)$ denotes the feature extractor. }
      \vspace{-5mm}
\end{figure*}

With a focus on the domain shift problem, DCASE 2022 challenge Task 2 launched a new task for unsupervised ASD \cite{baseline22,2022TOP1,2022TOP3,2022Top6}. Its dataset has hierarchical metadata of machine type, section ID and attributes, as we illustrated in Fig.~\ref{fig:framework}(b). Each section ID refers to a subset of the data within a domain shift scenario under a machine type, and domain shifts result from the change of attributes, e.g., the machine's operation speed and the environmental noise level. The 1$^\text{st}$ ranked method in the challenge \cite{2022TOP1} adopts self-supervised classification with section IDs as labels for feature learning. On the other hand, our previous work achieved 3$^\text{rd}$ place \cite{2022TOP3} using attributes as labels, considering their effect on acoustic characteristics. However, only using either section IDs \cite{2022TOP1, KuroyanagiNU-HDL2022} or attributes \cite{2022TOP3} may not be sufficient to obtain features helpful for characterising domain shifts. 

Existing methods \cite{DengTHU2022, WilkinghoffFKIE2022} used both section IDs and attributes in a parallel way. Taking attributes and section IDs in parallel assumes that the attributes and section IDs are independent or that the same attribute works equally under different domain shifts. However, the same attribute can impact the machine sound differently under different domain shift types (section IDs).
Thus, the relation between attributes and domain shift types has yet to be fully utilised despite their potential benefits for characterising domain shifts. 

This paper is the first study exploiting the implicit hierarchical relation between domain shift type and attribute for more effective feature learning for anomalous sound detection under domain shift. We propose a hierarchical metadata information constrained (HMIC) self-supervised method using domain shift types and attributes in a hierarchical way. Specifically, we set the attribute groups (AGs) under each section ID (domain shift type) to cluster the data with the same attributes' values as an attribute group, as shown in Fig.~\ref{fig:framework} (b). Then, we use the hierarchical relation as the constraint in self-supervised learning to obtain finer audio feature representation, with the section IDs characterising the type of domain shift for low-level feature learning and the attribute groups exploiting acoustic characteristics of each domain shift for high-level feature learning. Moreover, we propose an attribute group centre (AGC) based method to calculate anomaly scores. AGC represents the average of the learnt audio features from each attribute group. We calculate the anomaly score using the minimal Mahalanobis distance between the test sound's audio feature and AGCs to better adapt to the variance of domain shifts. Experiments conducted on the DCASE 2022 challenge Task 2 dataset demonstrate the proposed method's improved self-supervised audio feature learning compared to the state-of-the-art methods.

\section{PROPOSED METHOD}
\label{sec:2}

This section introduces our proposed HMIC, as illustrated in Fig.~\ref{fig:framework}, which consists of a backbone (i.e., MobileFaceNet \cite{MFN}) for feature extraction, and two classifiers (ID classifier $\mathcal{C}_{ID}$ and attribute classifier $\mathcal{C}_{AG}$) to predict section ID and attribute group label, respectively. It uses a hierarchical metadata information structure to exploit the implicit relation between section ID and attributes for finer feature learning. In addition, we introduce an AGC-based method for calculating the anomaly scores in the detection stage. 

\subsection{Hierarchical Metadata Information Structure}
\label{sec:2_1}

Addressing domain shift in ASD, metadata information (i.e., section IDs and attributes) related to domain shift is utilised, with their hierarchical relation being further exploited through a hierarchical metadata information structure in Fig.~\ref{fig:framework} (b). To emphasise audio clips under each section ID may have certain attributes with different values, we cluster data with the same attributes' values as an attribute group under this section ID. Therefore, each section ID has several AGs, and each AG under the same section ID has the same attribute types but different values. So, we constructed a metadata information tree for each machine type, with section IDs as nodes and AGs as leaves, as in Fig.~\ref{fig:framework} (b). 

Take the machine type ToyCar in DCASE 2022
challenge Task 2 \cite{baseline22} as an example, section ID\_00 contains four attributes (i.e., ``car model", ``speed", ``microphone number", and ``noise number") with different attribute values. ``car model" has the value of A1, C2, etc., and ``noise number" has the value of 1, 2, etc. By grouping these attributes corresponding to their values, we obtain 11 AGs for section {ID\_00}, and a total number of 44 AGs for the machine type ToyCar. Thus, the DCASE 2022 dataset of 7 machine types each with 6 section IDs, becomes 250 AGs under 42 section IDs to construct the hierarchical relation between section ID and attributes. Therefore, we can employ this hierarchical relation between section IDs and attributes as the constraint to learn finer audio features to mitigate the domain shift issue in ASD.

\subsection{Hierarchical Constrained Classification}
\label{sec:2_2}
With the hierarchical information discussed above, we can employ section IDs and AGs as the self-supervision labels to constrain the learning of the low-level and high-level audio features, respectively, as shown in Fig. \ref{fig:framework} (a). Here, low and high levels indicate the output coarse and fine-grained features from our model's low and high-level layers with the hierarchical constraint, respectively.

The log-Mel spectrogram ${\textbf{\textit{X}}}$ of the input audio signal $\textbf{\textit{s}}$ is the input of our model. We obtain the low-level feature $\textbf{\textit{f}}_{l}$ and high-level feature $\textbf{\textit{f}}_{h}$ from the backbone network (MobileFaceNet \cite{MFN}) via a feature extractor $\mathcal{F}(\cdot)$ and a 2-dimensional convolutional layer (Conv2D), respectively,
\begin{equation}
\label{eq:1}
    \textbf{\textit{f}}_{l}= \mathcal{F}(\textbf{\textit{X}})
\end{equation}
\begin{equation}
\label{eq:2}
    \textbf{\textit{f}}_{h} = \text{Conv2D}(\textbf{\textit{f}}_{l})
\end{equation}
To utilise hierarchical relation to learn features relevant to domain shift, section ID and AG are employed as self-supervision labels, $l_{ID}$ and $l_{AG}$, to constrain the learning of $\textbf{\textit{f}}_{l}$ and $\textbf{\textit{f}}_{h}$, respectively. 
First, two simple linear classifiers (ID classifier $\mathcal{C}_{ID}$ and attribute classifier $\mathcal{C}_{AG}$) are adopted for section ID label and AG label prediction, that  $\hat{l}_{ID} = \mathcal{C}_{ID}(\textbf{\textit{f}}_{l})$ and $\hat{l}_{AG} = \mathcal{C}_{AG}(\textbf{\textit{f}}_{h})$, respectively.  Then, an ID loss $\mathcal{L}_{ID}$ and an attribute loss $\mathcal{L}_{AG}$ are introduced to constrain the process of learning the low-level and high-level features, using e.g. the cross-entropy loss (\textit{CE}), 
\begin{equation}
\label{Eq:Loss_ID}
    \mathcal{L}_{ID} = CE(l_{ID}, \hat{l}_{ID})
\end{equation}
\begin{equation}
\label{Eq:Loss_AG}
    \mathcal{L}_{AG} = CE(l_{AG}, \hat{l}_{AG})
\end{equation}
Finally, the total loss $\mathcal{L}{_{total}}$ for model training is 
\begin{equation}
\label{Eqn:totalloss}
    \mathcal{L}{_{total}} = \lambda \mathcal{L}_{ID} + (1-\lambda)  \mathcal{L}_{AG}
\end{equation}
The weight $\lambda$ is empirically tuned for each machine type.

\subsection{Attribute Group Centre based Anomaly Detection}
\label{sec:2_3}
Anomaly score calculation is the key to evaluating the test sound in the anomaly detection stage. We introduce the attribute group centre (AGC) to calculate the anomaly score. Each attribute group's AGC is the average of the learnt audio features from that attribute group. Then, the feature of the test sound is compared with all the AGCs to allow better anomaly detection in the presence of domain shift. 

Assume $N$ training audio clips under the $m$-th attribute group with the label ${l_{AG}}_{m-1}$, $m = 1, 2, \dots, M$ and $M$ is the number of attribute groups under corresponding section ID. The $m$-th attribute group centre $\textbf{\textit{c}}_{m}$ is 
\begin{equation}
\label{eq:6}
    \textbf{\textit{c}}_{m} =\frac{1}{N} \sum_{n=1}^{N}{\hat{\textbf{\textit{f}}}_{h_{n}}}
\end{equation}
where ${\hat{\textbf{\textit{f}}}_{h_{n}}}$ denotes the high-level audio feature derived from the model for the $n$-th training audio clip, $n = 1, 2, \dots, N$. 

Then, Mahalanobis distance \cite{MD} is used to measure similarity between the audio feature representation $\overline{\textbf{\textit{f}}}$ of the sound under test and each AGC  $\textbf{\textit{c}}_{m}$, $m = 1, 2, \dots, M$, and the minimal Mahalanobis distance is taken as the anomaly score $\mathcal{A}$
\begin{equation}
\label{frm:md}
    \mathcal{A} = \underset{m\in [1, M]}{\min}\sqrt{(\overline{\textbf{\textit{f}}} - \textbf{\textit{c}}_{m})^{T} {\mathbf{\Sigma}^{-1}}(\overline{\textbf{\textit{f}}} - \textbf{\textit{c}}_{m})}
\end{equation}
where $\mathbf{\Sigma}^{-1}$ is the inverse of the covariance matrix $\mathbf{\Sigma}$, and $\mathbf{\Sigma}$ is obtained from the feature of all the audio clips under the $m$-th attribute group of the same section. 

Our proposed ASD method with the AGC-based anomaly score calculation is named \textbf{HMIC-AGC}. The proposed method is later compared with \textbf{HMIC-DC}, which calculates the Mahalanobis distance between $\overline{\textbf{\textit{f}}}$ and the domain centre (DC), i.e., the average feature of each domain, instead of each attribute group, to derive the anomaly score. Though DC is widely used for anomaly score calculation, such as the 1$^\text{st}$ ranked method in the DCASE 2022 challenge \cite{2022TOP1}, DC uses average features from multiple domains, while AGC considers the acoustic characteristics of each specific domain to have more accurate feature representation under domain shifts.

\section{EXPERIMENTAL RESULTS}
  
\subsection{Experimental Setup}

\noindent\textbf{\textit{Dataset}}
The training data for model training is from the development and the additional datasets of the DCASE 2022 challenge Task 2 \cite{baseline22}, which includes five different machine types (bearing, fan, gearbox, slider, and valve)  \cite{MIMIIDG} and two types of toys (i.e., ToyCar and ToyTrain) \cite{ToyADMOS2}. Each machine type contains six section IDs, each with 990 and 10 audio clips from the source and target domain, respectively. We evaluate the performance on the evaluation dataset of the DCASE 2022 challenge Task 2. Note that the evaluation dataset's domain information (source and target) is unknown to verify the generalization ability of the ASD systems.
\begin{table*}[htbp]
\centering
\caption{Performance comparison in terms of AUC (\%) and pAUC (\%) on the evaluation dataset of DCASE 2022 challenge Task 2. \textbf{Total}: harmonic mean (\%) of AUC and pAUC scores over all the machine types, sections, and domains.
}
\label{tab:1}
\resizebox{0.98\textwidth}{!}{
\begin{tabular}{@{}ccccccccccccccccc@{}}
\toprule
\multirow{2}{*}{\textbf{Methods}} &
  \multicolumn{2}{c}{\textbf{ToyCar}} &
  \multicolumn{2}{c}{\textbf{ToyTrain}} &
  \multicolumn{2}{c}{\textbf{Bearing}} &
  \multicolumn{2}{c}{\textbf{Fan}} &
  \multicolumn{2}{c}{\textbf{Gearbox}} &
  \multicolumn{2}{c}{\textbf{Slider}} &
  \multicolumn{2}{c}{\textbf{Valve}} &
  \multicolumn{2}{c}{\textbf{Total}} \\ 
    \cmidrule(lr){2-3} \cmidrule(lr){4-5} \cmidrule(lr){6-7} \cmidrule(lr){8-9} \cmidrule(lr){10-11} \cmidrule(lr){12-13} \cmidrule(lr){14-15} \cmidrule(lr){16-17}
                            & AUC   & pAUC  & AUC   & pAUC  & AUC   & pAUC  & AUC   & pAUC  & AUC   & pAUC  & AUC   & pAUC  & AUC   & pAUC  & AUC   & pAUC  \\ \midrule
AE \cite{AEMNV2}            & 61.18 & 60.21 & 43.14 & 49.36 & 59.93 & 53.95 & 41.16 & 50.12 & 61.92 & 51.95 & 58.95 & 54.16 & 54.26 & 51.30 & 53.01 & 52.80 \\
MobileNetV2 \cite{AEMNV2}   & 42.79 & 53.44 & 51.22 & 50.98 & 58.23 & 52.16 & 50.34 & \textbf{55.22} & 51.34 & 48.49 & 62.42 & 53.07 & 72.77 & 65.16 & 54.19 & 53.67 \\
Attribute-only \cite{2022TOP3}        & 87.61 & 73.12 & 56.64 & 52.60 & \textbf{73.92} & 58.77 & 52.69 & 49.79 & 74.11 & 59.96 & 73.39 & 59.51 & 78.14 & 69.26 & 67.68 & 59.47 \\ 
Domain-only \cite{2022TOP1}        & 77.15 & 67.47 & 55.92 & 51.53 & 71.91 & \textbf{60.74} & 54.52 & 53.86 & 78.75 & 53.30 & 78.87 & \textbf{59.56} & 85.60 & 78.59 & 69.51 & 59.56 \\\midrule
\textbf{HMIC-DC}               & 82.44 & 71.92 & 57.88 & 52.75 & 67.45 & 59.14 & 56.55 & 53.03 & 77.22 & 59.74 & 80.59 & 58.75 & 89.70 & \textbf{82.69} & 70.20 & 61.15\\
\textbf{HMIC-AGC}     & \textbf{87.91} & \textbf{77.51} & \textbf{59.10} & \textbf{52.83} & {68.14} & {59.41} & \textbf{57.63} & 53.25 & \textbf{79.78} & \textbf{61.29} & \textbf{80.76} & 58.29 & \textbf{89.87} & 82.30 & \textbf{71.79} & \textbf{61.91} \\ \bottomrule\bottomrule
\end{tabular}}
\vspace{-3mm}
\end{table*}

\noindent\textbf{\textit{Evaluation Metrics}} 
The evaluation metrics include the area under the receiver operating characteristic curve (AUC), partial AUC (pAUC), and the harmonic mean of AUC and pAUC scores over all the machine types, sections, and domains \cite{baseline22}.

\noindent\textbf{\textit{Implementation}}
The log-Mel spectrogram of the audio clips is used as the input feature for our model, where the frame size is set as 1024 with an overlap of 50\%, and the number of Mel filter banks is 128. The dimension of the input log-Mel spectrogram is 128 × 313. Our model is trained with 120 epochs, using Adam optimizer \cite{Adam} with a learning rate of 0.0001, and the cosine annealing strategy is applied for learning rate decay. 

\subsection{Experimental Results}
\begin{figure}[!t]
 \centering
 \subfloat[\centering{
 Domain-only
 \cite{2022TOP1}}]{
 \includegraphics[width=0.9\columnwidth]{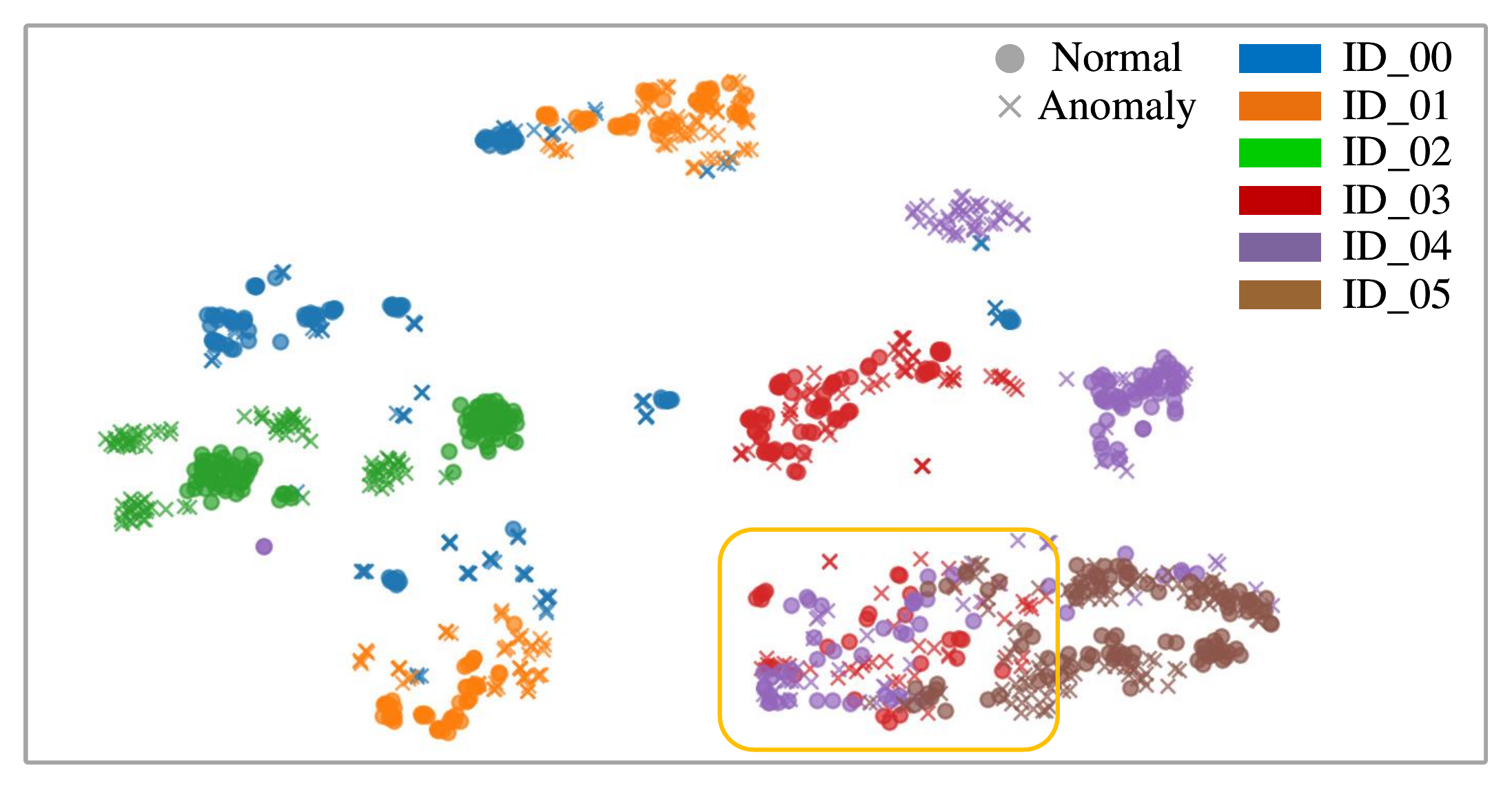}
     }
 \vspace{-3mm}
 \quad
 \subfloat[\centering{
 Attribute-only 
 \cite{2022TOP3}}]{
 \includegraphics[width=0.9\columnwidth]{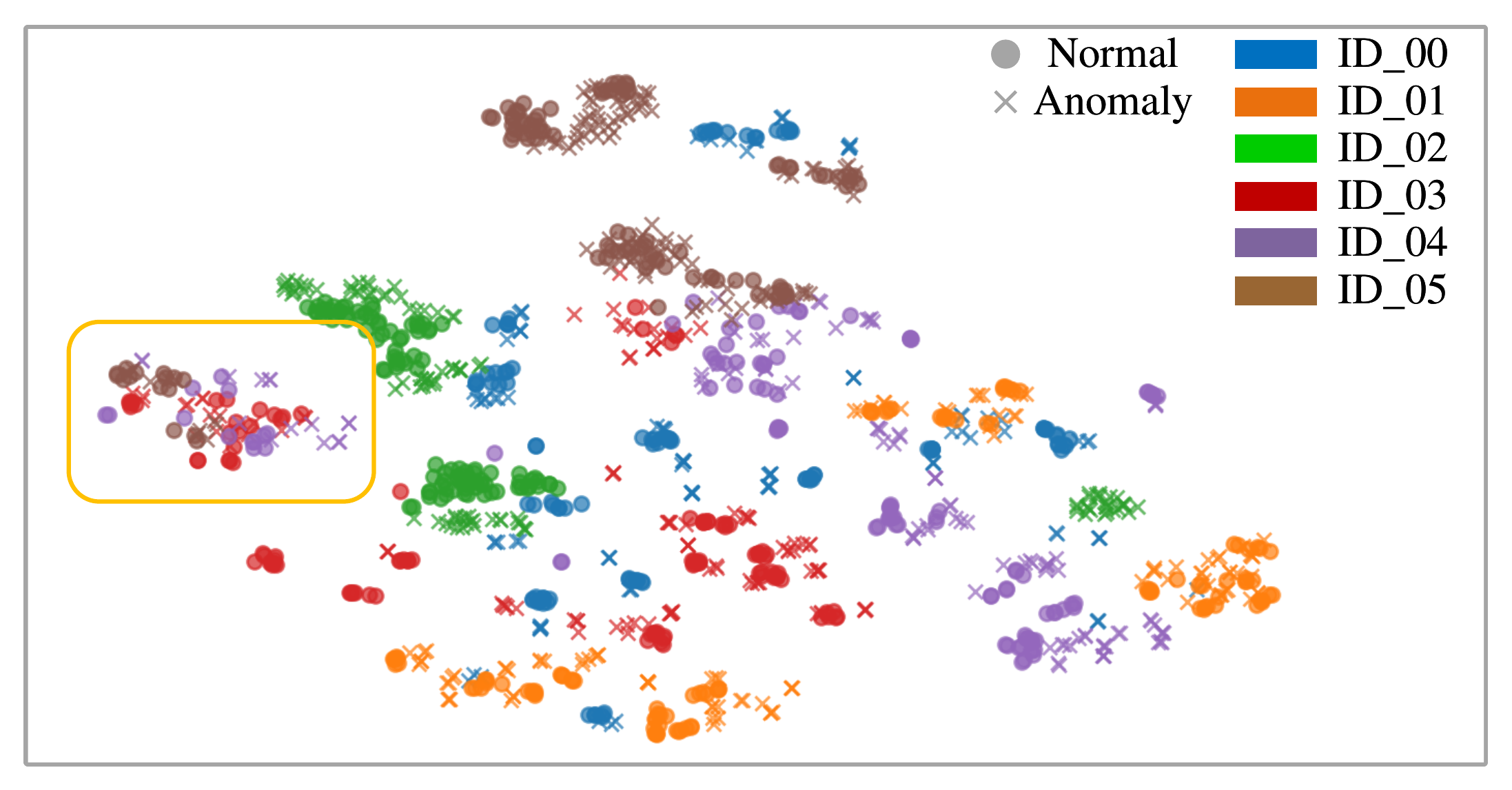}
     }
 \vspace{-3mm}
 \quad
 \subfloat[\centering{
 HMIC (proposed)
 }]{\includegraphics[width=0.9\columnwidth]{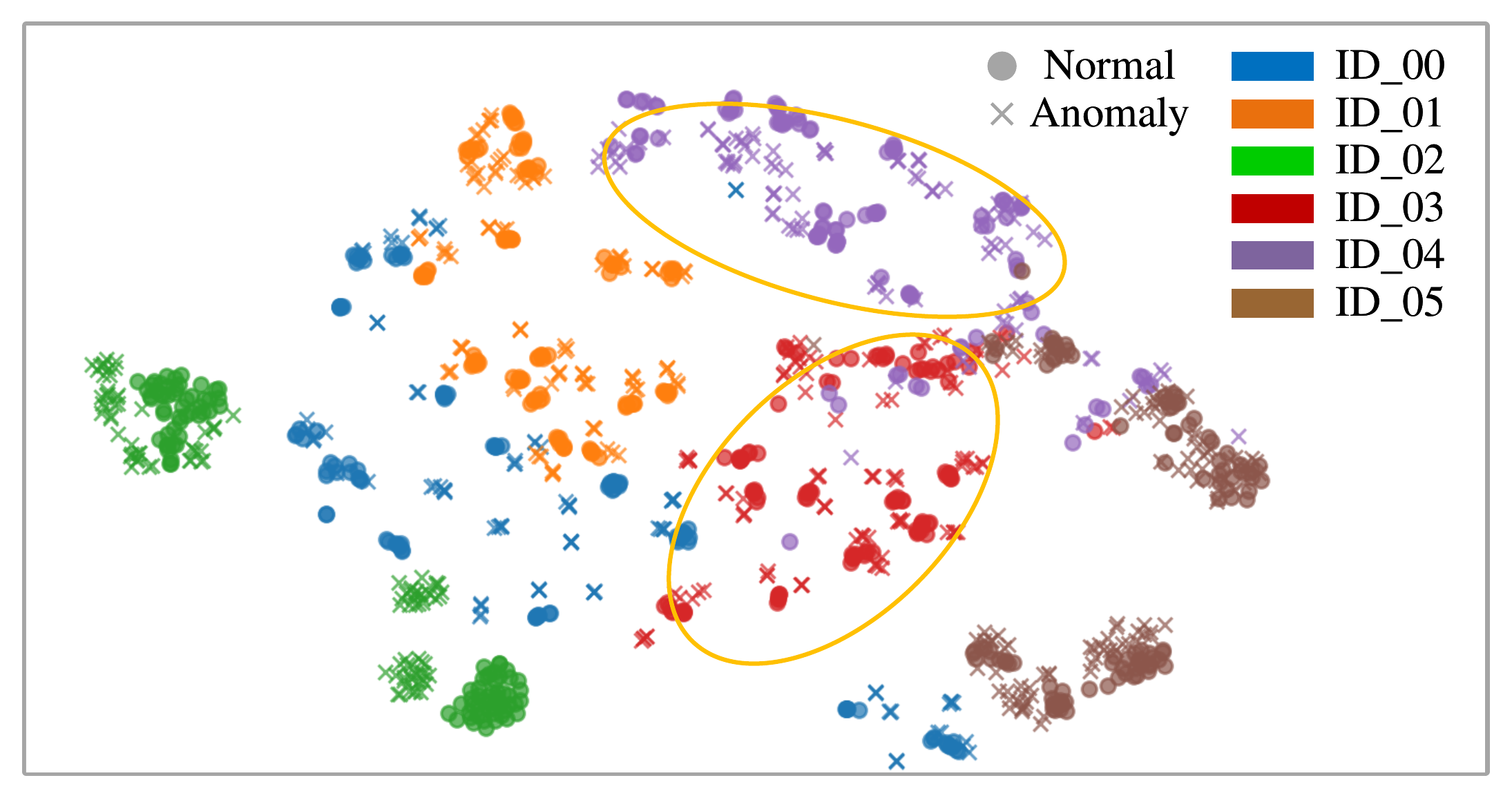}
     }
 \caption{The t-SNE visualisation of the learnt audio features using different self-supervised methods for machine type Bearing. Different colours represent different section IDs. ``•'' and ``×'' respectively represent normal and anomalous sounds.}
 \label{fig:tsne}
 \vspace{-3mm}
\end{figure}
\noindent\textbf{\textit{Performance Comparison}} 
Our proposed HMIC-AGC method hierarchically uses the domain shift type and attribute for self-supervised ASD with AGC for anomaly score calculation. In the experiment, it was compared with baseline methods (AE \cite{AEMNV2} and MoblieNetV2 \cite{AEMNV2}), the self-supervised methods only using the attribute (Attribute-only, as the 3$^\text{rd}$ ranked method in DCASE 2022 Task 2 \cite{2022TOP3}) or section ID (Domain-only, as the 1$^\text{st}$ ranked method \cite{2022TOP1}), and HMIC-DC using DC for anomaly score calculation. For a fair comparison, all methods are performed without model pretraining, and all adopt log-Mel spectrogram as the input without the high-pass filter in \cite{2022TOP1} or temporal information fusion in \cite{2022TOP3}.

As can be seen from Table~\ref{tab:1}, both HMIC-DC and HMIC-AGC can significantly improve the detection performance for all machine types except Fan, as compared with the baseline methods, i.e., AE \cite{AEMNV2} and MoblieNetV2 \cite{AEMNV2}. In addition, the proposed methods achieve the best overall performance compared to Domain-only or Attribute-only methods from the 1$^\text{st}$ and 3$^\text{rd}$ ranked submissions in DCASE 2022, respectively. Although the pAUC performance on Slider and Fan of HMIC-AGC is slightly lower than the Domain-only method, it significantly improves the pAUC performance on ToyCar and Gearbox, with 10.04\% and 7.99\% improvement, respectively. Specifically, both HMIC-DC and HMIC-AGC can improve the total harmonic mean performance, and HMIC-AGC achieves the best total harmonic mean performance. The results demonstrate the effectiveness of the hierarchical metadata information constraint and the AGC-based anomaly scores calculation, which show the superior generalisation ability of our proposed methods for ASD under domain shift conditions. In addition,  the performance on the development set has the same trend as that on the evaluation set, though not presented in this paper.

\noindent\textbf{\textit{Visualisation Analysis}} 
To further verify the effectiveness of the proposed HMIC for improved feature learning, the test data (with all 6 section IDs) from both development and evaluation datasets are evaluated. The t-distributed stochastic neighbour embedding (t-SNE) \cite{tSNE} cluster visualisation of the learnt features using section ID only, attribute only, or the proposed method are illustrated in Fig.~\ref{fig:tsne}. It can be seen that audio features are misclassified in the presence of overlapping between sections ID\_03 and ID\_04, when only using section ID or attribute (metadata without hierarchical relation) for self-supervised classification. In contrast, they can be distinguished with the proposed method from different sections as the areas marked with orange in Fig.~\ref{fig:tsne} (c). The results demonstrate the effect of HMIC for more distinguishable feature learning under domain shift.

\section{CONCLUSION}
\label{sec:4}
We have presented a self-supervised method for anomalous sound detection under domain shift, where a hierarchical metadata information structure is constructed and used as the constraint in self-supervised learning for improved feature learning. In addition, an attribute group centre based anomaly scores calculation method is introduced, which further enhances the domain generalisation ability by considering the attributes of domain shift. Experimental results demonstrate the effectiveness of the proposed method, with substantial improvements in the audio feature learning over those that only use section ID or attributes, as in the state-of-the-art methods in DCASE 2022 challenge Task 2. 

%
\bibliographystyle{IEEEtran}

\bibliography{refs}

\end{document}